\documentclass[11pt, a4paper]{article}

\usepackage[utf8]{inputenc}
\usepackage[T1]{fontenc}
\usepackage{geometry}
\usepackage{amsmath,amssymb,amsfonts}
\usepackage{algorithm,algpseudocode}
\usepackage{graphicx}
\usepackage{textcomp}
\usepackage{multirow}
\usepackage{hyperref}
\usepackage{authblk} 
\usepackage{fancyhdr} 
\usepackage[numbers,sort&compress]{natbib} 

\newcommand{\copyrightnotice}{%
  \thanks{\copyright~2023. This is the author's version of a work that was accepted for publication in \textit{Energy} following peer review. 
  Changes resulting from the publishing process, such as structural formatting and other quality control mechanisms, may not be reflected in this document. 
  The definitive version of record is available at DOI: \href{https://doi.org/10.1016/j.energy.2023.130110}{10.1016/j.energy.2023.130110}}
}

\usepackage{nomencl}
\makenomenclature

\def\lowcomma{_{\textstyle,}}
\def\lowperiod{_{\textstyle.}}

\title{\textbf{Blockchain-Enabled Renewable Energy Certificate Trading: A Secure and Privacy-Preserving Approach}\copyrightnotice}

\author[1]{Wei-Jen Liu}
\author[1]{Wei-Yu Chiu\thanks{Corresponding author: chiuweiyu@gmail.com}}
\author[2]{Weiqi Hua}

\affil[1]{\small Department of Electrical Engineering, National Tsing Hua University, Hsinchu 300044, Taiwan}
\affil[2]{\small School of Engineering, University of Birmingham, Birmingham B15 2TT, UK}

\date{} 

\begin{document}

\maketitle

\begin{abstract}
In the 21st century, transitioning to renewable energy sources is imperative, with fossil fuel reserves depleting rapidly and recognizing critical environmental issues such as climate change, air pollution, water pollution, and habitat destruction. Embracing renewable energy is not only an environmental necessity but also a strategic move with multiple benefits. By shifting to renewable energy sources and supporting their production through the acquisition of renewable energy certificates, we foster innovation and drive economic growth in the renewable energy sector. This, in turn, reduces greenhouse gas emissions, aligning with global efforts to mitigate climate change. Additionally, renewable energy certificates ensure compliance with regulations that mandate the use of renewable energy, enhancing legal adherence while promoting transparency and trust in energy sourcing. 

To monitor the uptake of renewable energy, governments have implemented Renewable Energy Certificates (RECs) as a tracking mechanism for the production and consumption of renewable energy. However, there are two main challenges to the existing REC schema: 1) The RECs have not been globally adopted due to inconsistent design; 2) The consumer privacy has not been well incorporated in the design of blockchain. In this study, we investigate the trading of RECs between suppliers and consumers using the directed acyclic graph (DAG) blockchain system and introduce a trading schema to help protect consumer information. Our results demonstrate lower transaction time by 41\% and energy consumption by 65\% compared to proof-of-stake.
\end{abstract}

\vspace{1em}

\printnomenclature
\nomenclature{$h$}{Time slot $h$ where $t=1,2,...,H$.}
\nomenclature{$i$}{Supplier $i$ where $i=1,2,...,I$.}
\nomenclature{$j$}{Consumer $j$ where $j=1,2,...,J$.}
\nomenclature{$p^{\text{max}}_h$}{Maximum REC selling price in the current market in time slot $h$.}
\nomenclature{$t^{\text{max}}$}{Max time period to buy RECs.}
\nomenclature{$t^{\text{remain}}$}{Remaining time period to buy RECs.}
\nomenclature{$C_{j, h}$}{REC bidding policy of consumer $j$ in time slot $h$.}
\nomenclature{$D_{j,h}$}{Penalty fee that the consumer $j$ incurs for not meeting the green ratio in time slot $h$.}
\nomenclature{$E^{\text{res}}_{i, h}$}{Renewable energy generated by supplier $i$ in time slot $h$.}
\nomenclature{$E^{\text{res}}_{j, h}$}{Renewable energy generated by consumer $j$ or bought from the main utility in time slot $h$.}
\nomenclature{$E^{\text{c}}_{j, h}$}{Total energy consumption by consumer $j$ in time slot $h$.}
\nomenclature{$G_{j,h}$}{Green ratio of consumer $j$ in time slot $h$.}
\nomenclature{$G^{\text{lag}}_{j,h}$}{Green ratio lag of consumer $j$ in time slot $h$.}
\nomenclature{$G^{\text{target}}_{j,h}$}{Green ratio target of consumer $j$ in time slot $h$.}
\nomenclature{$R^{\text{gen}}_{i, h}$}{Renewable energy certificate generated by supplier $i$ in time slot $h$.}
\nomenclature{$R^{\text{own}}_{j, h}$}{Number of RECs that consumer $j$ owns in time slot $h$.}
\nomenclature{$b_{i, h}$}{Initial purchase price of supplier $i$ in time slot $h$.}
\nomenclature{$P_{i, h}(b_{i, h})$}{Selling price policy of supplier $i$ in time slot $h$.}
\nomenclature{$Q_{j, h}$}{Quantity of REC to buy for consumer $j$ in time slot $h$.}
\nomenclature{$Q^{\text{max}}_{j, h}$}{Maximum quantity permitted for consumer $j$ to buy in time slot $h$.}
\nomenclature{$\alpha_{j,h}$}{Urgency of purchase by consumer $j$ in time slot $h$.}
\nomenclature{$\tau_{\text{remain}}$}{Remaining lifetime of REC.}
\nomenclature{$\tau_{\text{lifetime}}$}{Total lifetime of REC.}
\nomenclature{$\gamma$}{Unmet capacity coefficient.}
    
\section{Introduction}
\label{sec:Introduction}


As renewable energy consumption continues to rise \cite{16bhattacharya}, the topic of renewable energy trading has gained significant importance. According to the renewable portfolio standard \cite{01berry}, a Renewable Energy Certificate (REC) signifies one megawatt-hour of renewable energy generated by a certified renewable energy generator. This certificate is tradeable and validated by a designated organization or government. Each REC has a defined lifespan and will be retired upon expiration. If necessary, a new REC can be acquired upon the expiry of the old one. The advent of the World Wide Web \cite{12aghaei} has revolutionized REC trading, enabling transactions across international borders, among companies, and even households. RECs are available in two primary forms: bundled and unbundled \cite{13tanaka}. An unbundled REC involves the sale of proof of energy generated from renewable sources; a bundled REC entails the sale of both the REC and the corresponding renewable energy. Typically originating from newly established facilities, bundled RECs allow buyers to claim associated energy savings. However, they face limitations such as geographic constraints, energy transfer costs, and power grid infrastructure expenses, resulting in lower market liquidity but higher costs than unbundled RECs. Furthermore, some places are restrained from generating RECs due to geographic limits for generating renewable energy, which can cause high prices for a single REC. Generally, unbundled RECs are cost-effective, geographically flexible, and more liquid in the market.

To illustrate the practical implications of our research, let us consider a concrete scenario. In Taiwan, the government hosts T-REC \cite{22T-REC}, and the Center for Resource Solutions hosts Green-e \cite{22green-e}, both serving as online marketplaces. These platforms cater to companies that either do not invest in renewable energy or lack sufficient RECs. Companies can trade RECs in both bundled and unbundled forms. Unbundled RECs offer flexibility and facilitate a higher volume of transactions, but they necessitate meticulous REC traceability to prevent double-accounting \cite{08chen}. For instance, when a consumer purchases unbundled RECs, only the new owner can claim the use of renewable energy associated with those certificates, ensuring transparency and accuracy in tracking renewable energy consumption. Researches \cite{18noor, 21Paudel} have shown that an automated smart contracts can reduce the level of human resources required, and the blockchain's traceability and tamper-proof benefits are suitable for energy trading.

In the subsequent sections, we will delve into these platforms, analyzing the characteristics of bundled and unbundled forms of renewable energy certificate trading and their impact on market liquidity and cost. We will then discuss existing challenges and issues, focusing on how smart contracts based on blockchain can offer solutions for REC transactions, along with considerations for privacy protection to ensure the security of personal data during REC transactions.
Through this research, we provide insights into the development of the REC trading market, promoting wider adoption of renewable energy while ensuring transparency, security, and privacy in transactions. The objectives of this paper can be summarized as follows:
\begin{itemize}
    \item Discuss the challenges and issues associated with REC trading, particularly the potential for double-accounting, and propose solutions leveraging smart contracts based on blockchain.
    
    \item Address privacy concerns in REC transactions and suggest measures to ensure the security of personal data.
    
    \item Provide insights into the REC trading market, promoting wider adoption of renewable energy while emphasizing transparency, security, and privacy in transactions.
\end{itemize} 

The conventional blockchain structure suffers from scalability and validator aggregation problems related to the blockchain ledger size and transaction performance. Furthermore, because the transaction data is recorded on the blockchain and can be accessed by anyone, although the Identity data is encrypted, it can still reveal some critical information for commercial building consumers by analysing large enough data, such as trading pattern or address usually used. Meanwhile, the European Union proposed General Data Protection Regulation (GDPR) \cite{17voigt} and claims that data controllers must take technical and organizational measures to minimize personal data leakage. Therefore, privacy preservation is essential for all consumers. Lin \textit{et al.} \cite{20lin} designed a blockchain architecture to protect user privacy based on the Ethereum and account models. It uses group signatures and broadcast encryption to shield data from other users.

Sharding chain \cite{18zamani}, a popular research method for scalability and validator aggregation problems, enables partitioning the computational and storage workload across a peer-to-peer network \cite{20khalid, 18wang} to reduce the entire network's load. Nevertheless, there are still many challenges for the sharding chain. First, the sharding chain is much smaller than the main chain, and a single shard as these network size is usually much smaller. Therefore it is easier to take over by malicious nodes. Second, cross-shard communication becomes complex when data is synchronously requested from multiple shards. Another popular method is off-chain transactions \cite{20Al, 17imbault}, a more disruptive approach to address scalability and validator aggregation problems. Off-chain transactions allow transactions outside the network and more effectively improve speed and reduce throughput limitations than on-chain. After the transaction, the complete results must be validated on the blockchain network. Nonetheless, off-chain transactions can be susceptible to fluid fake-data attacks. The other research, named prism \cite{19bagaria} proposed by Bagaria \textit{et al.}, is a new proof-of-work blockchain protocol that aims to improve the original blockchain design. The design is based on deconstructing Nakamoto's blockchain and scaling it up. It can achieve optimal transaction throughput and provide strict consistency yet sacrifice scalability and energy consumption as a tradeoff \cite{20wang}.

Sharding chain and off-chain transactions are newly research topic to addressing the scalability and validator aggregation problem; However, research and PoW-based networks still have some cons that need to be resolved. Another out-of-frame solution that has been investigated by Park \textit{et al.}, uses directed acyclic graphs (DAGs) \cite{19park}. The DAG, a new generation of blockchain, is a graph in which a node has a direction and points to other nodes. DAG is suitable for solving the scalability problem for two main reasons: First, it is not limited by the time it takes to create a new block and can handle more transactions per second than typical blockchains. Second, it is validated by its neighbor nodes through using a voting mechanism with majority rule \cite{18conti} without requiring miners, thus reducing energy consumption, avoiding high transaction fees, and preventing front-running attacks \cite{19eskandari, 19daian}, and it has better decentralization. Therefore, the DAG has strong potential and deserves further research.

Motivated by the advantages of DAG over conventional blockchain, we propose a DAG-based REC trading platform. Because the DAG blockchain concept has emerged recently, the application of DAG-based blockchain to REC trading has not yet been investigated. A smart meter helps analyze renewable energy consumption usage and REC trading online in our scenario. The DAG network structure also solves the scalability and validator aggregation problem, because it does not separate into a different chain like shard chain and off-chain transactions, it prevents cross-shard communication problems also not easy to take over the entity blockchain network because the blockchain ledger size is larger than other separated chains. However, the consensus algorithm commonly used on DAG, the standard voting-based fast probability consensus (FPC), found a weakness \cite{19capossele} caused by the randomness of threshold can be guessable.

In this paper, to address the blockchain scalability problem and privacy-preserving on REC trading, 1) we applied the FPC with a reputation concept, such as trading activity or assets, as the weight for the voting mechanism to verify transactions. The asset is determined by the organization's scale and is related to its reputation. We implemented the proposed procedure for trading RECs based on supply and demand for commercial building consumers. Each trader is assigned with a reputation value intended to encourage users and enhance FPC on the REC trading platform, further preventing malicious attacks due to the random weakness of the consensus algorithm. 2) We propose a privacy-preservation algorithm to address these privacy challenges in the trading phase. Creating different types of accounts according to the user's activity and transaction volume can maintain the fairness of trading and protect privacy of consumers. 3) Finally, we conducted numerical analyses with real-world data, and a comprehensive model was constructed to simulate REC trading. This analysis revealed that the proposed trading platform outperforms the standard blockchain structure, such as bitcoin and ethereum, which means blocks are linked together on a single chain. The scalability and validation performance are significant when the blockchain ledger size increases and creates more liquidity in the REC market. Furthermore, the proposed method also reveals better decentralization among PoW and PoS consensuses.

The proposed REC trading system utilizes innovative techniques, including directed acyclic graph (DAG) blockchain, voting-based fast probability consensus, and abstract accounts. These techniques offer several advantages over traditional blockchain-based REC trading systems, including:
      \begin{itemize}
          \item Improved performance and scalability: DAG blockchains are more efficient and scalable than traditional blockchains, resulting in faster transaction times and lower energy consumption. This is important for a REC trading system, as it allows buyers and sellers to quickly and easily trade RECs without sacrificing performance.
          \item Enhanced security and privacy: Abstract accounts protect users' privacy in the REC trading system, while the voting-based fast probability consensus algorithm makes the system more secure and reliable. This is important to ensure that the REC trading system is fair and trustworthy for all participants.
          \item Global adoption: The proposed REC trading system is designed to be globally adopted and promote renewable energy use. This is achieved by using a DAG blockchain, which is more efficient and scalable than traditional blockchains, and by using abstract accounts, which protect users' privacy.
        \end{itemize}
      In addition to these benefits, the proposed REC trading system is also designed to be user-friendly and easy to use. This makes it accessible to many participants, including individuals, businesses, and governments.
      
This study presents contributions at the intersection of renewable energy trading, blockchain technology, and privacy preservation. Firstly, our research focuses on the Renewable Energy Certificate (REC) trading market using a Directed Acyclic Graph (DAG) blockchain. This innovative approach substantially reduces computational demands and power consumption during block creation compared to traditional blockchain structures. Additionally, we propose a reputation-based consensus mechanism within the DAG for REC trading, a concept not explored in existing literature, enhancing transaction validation speed and market scalability. Secondly, we developed a comprehensive privacy-preserving transaction schema for REC trading. This schema ensures sensitive user information remains protected by integrating privacy features into transaction elements such as bids, offers, and user REC holdings, thus encouraging broader participation in the renewable energy trading market. Beyond technological advancements, we emphasize the crucial aspect of energy conservation. By streamlining REC trading, our approach indirectly promotes energy conservation efforts, aligning with global sustainability goals by encouraging the adoption of renewable energy sources and reducing reliance on fossil fuels. Lastly, our research holds valuable policy implications by providing insights for policymakers and market regulators to design and implement effective policies promoting renewable energy adoption. Understanding the potential of DAG-based REC trading markets and the significance of privacy preservation allows for the creation of an enabling policy environment that accelerates the transition to clean energy, fostering sustainable energy ecosystems. By amalgamating technological innovation, privacy preservation, energy conservation, and policy insights, this paper advocates for a comprehensive approach to renewable energy trading.

The remainder of the paper is organized as follows. Section \ref{sec:model} discusses the system model with three different types of users and its interactive modelling to the decentralized REC platforms. Section \ref{sec:approach} presents the blockchain-enabled platform in detail with the smart contract, consensus and privacy preserving schema design. Numerical results are presented in Section \ref{sec:result}. REC trading in the blockchain market is also discussed, and the numerical results regarding blockchain energy reduction and transaction performance improvement are compared with existing proposals. In addition, privacy preservation based on the proposed trading schema is also analyzed in this section.


\section{Related works}
\label{sec:related}
This section briefly discusses various approaches to REC on blockchain systems, covering different types of blockchain systems, privacy protection, and zero-knowledge proof verification. Blockchain-based solutions can provide a secure and efficient way to transact RECs. Consideration is given to preserving privacy and protecting the sensitive data of individual users and commercial buildings that transact RECs. Cryptographic techniques such as implementing zero-knowledge proofs are provided to verify the validity of transactions without revealing sensitive information.

        To understand green energy certificate trading, S. Zeng and A. Tenveer \cite{22modeling} provide a comprehensive review that explores critical factors influencing consumers' adoption of Green Energy Technologies (GETs) in Pakistan and sheds light on insights to drive green energy certificate trading. GETs, recognized for their environmentally friendly nature, hold significant promise in achieving net-zero carbon goals. Despite government initiatives, consumer adoption remains low, indicating underlying challenges. The review emphasizes enhancing awareness, policy reforms, cost-effectiveness, and user-friendliness as crucial elements. The findings offer valuable insights for stakeholders and experts, providing a foundation to encourage green energy certificate trading.
        
        A vital point of this research is using blockchain-based systems to facilitate near real-time energy transactions between prosumers and consumers without needing a central authority. Research by Li et al. \cite{18li} and Yao et al. \cite{19yao} emphasizes the application of blockchain-based systems for facilitating resource transactions within IIoT networks. Their studies advocate the utilization of decentralized ledgers, smart contracts, and encryption to ensure the security and privacy of these transactions. This approach holds the promise of enhancing traceability, accountability, and sustainability in energy and carbon markets. Additionally, Su et al. \cite{19su}, Wang et al. \cite{19wang}, and Hua et al. \cite{19hua} have directed their attention towards harnessing blockchain technology for the integration of renewable energy sources and electric vehicles (EVs) into smart grids and other IIoT networks. Their work proposes the implementation of blockchain-based systems to manage EV charging and carbon trading, highlighting the potential advantages in terms of traceability, accountability, and sustainability.

        Recent research examples by Baris \cite{12baris} have illustrated the challenges and potential benefits of integrating renewable energy into smart grids and other power systems, along with the role of government policies and incentives in facilitating their adoption. Study by Barlongo \cite{19barlongo} explored the potential of blockchain technology to enable efficient and secure energy transactions in smart grids and other decentralized energy systems using the Ethereum blockchain. The growth of self-consumption and local renewable energy communities, particularly in California, North Carolina, Arizona, and Nevada, has led to the need for efficient mechanisms to manage energy transactions between prosumers and consumers. Authors Khalid et al. \cite{20khalid} and Ali et al. \cite{20ali} have proposed using blockchain technology to handle and facilitate energy transactions. These studies highlight the potential benefits of using decentralized ledgers and smart contracts to manage energy transactions in local markets.
                
        The integration of blockchain technology in renewable energy markets has received significant attention in recent years. It has been used to improve the current renewable energy certificate trading market for governments and large institutions \cite{22T-REC, 22green-e}. Several studies have proposed blockchain-based renewable energy certificate (REC) management and trading systems to promote renewable energy adoption and ensure certificate validity through blockchain technology's traceability, accountability, and sustainability.

        Recent studies have explored the potential of using blockchain technology in the renewable energy market. For example, Castellanos et al. \cite{17castellanos} have shown that blockchain-based REC trading systems have the potential to reduce transaction costs, improve transparency and accountability in REC markets, and enable participation by small-scale energy consumers. Zhao et al. \cite{20zhao} proposed a blockchain-based personal REC called I-Green, which uses a generative proof consensus protocol to facilitate the adoption of distributed renewable energy. Kim et al. \cite{20kim} examined the use of sealed bid auctions combined with blockchain technology to rationalize transactions among market participants in renewable energy markets. Dimcho et al. \cite{20Dimcho} proposed a blockchain-based system that makes RECs efficient, trusted, and anonymous and evaluates their performance through implementation and testing. Knirsch et al. \cite{20knirsch} proposed a decentralized and license-free system for issuing and verifying RECs using the Gecko protocol. This system enables off-chain REC transactions and aims to address the challenges of decentralized and diverse energy markets. In addition, Hsiao et al. \cite{18hsiao} examined potential fraud and validation issues in RECs and proposed a blockchain-based system to eliminate these issues. As such, blockchain technology has been identified to improve transparency and accountability in REC markets.

        Previous work has extensively investigated the front-running issue in decentralized applications (DApps) deployed on blockchain platforms. Zamani et al. \cite{18zamani} proposed a sharding-based public blockchain protocol capable of processing over 7300 transactions per second with low confirmation latency. Additionally, Bagaria et al. \cite{19bagaria} introduced a new proof-of-work blockchain protocol designed for optimal throughput and confirmation latency. Building upon these prior studies, this paper presents a novel solution to mitigate front-running attacks in blockchain technology. Our approach leverages a combination of the consensus mechanism of Directed Acyclic Graphs (DAG) and reputation systems to enhance both the security and efficiency of blockchain operations. We conducted simulations to thoroughly assess the effectiveness of our proposed solution and provided a comparative analysis against existing approaches.
                
        The use of blockchain technology for privacy-preserving energy transactions has been widely investigated. Chaudhary et al. \cite{19chaudhary} proposed using blockchain-based systems to protect the confidentiality and integrity of energy transaction data. Gai et al. \cite{19gai} included the development of blockchain-based consortium systems to protect user privacy and the privacy of smart grids. Rong et al. \cite{20Rong} considered privacy protection when using smart meters for transactions with blockchain, using a federated chain approach and integrating ring encryption and decentralized data storage methods to achieve privacy protection. Aitzhan et al. \cite{16aitzhan} investigated  decentralized energy transactions facilitated by blockchain technology, multi-signature mechanisms, and secure, anonymous cryptographic communication. Lin et al. \cite{20lin} developed a privacy-preserving, permissioned blockchain architecture tailored for regulatory use cases. These studies highlighted the potential benefits of using cryptography, such as zero-knowledge proofs and secure multi-party computation, to ensure the security and privacy of energy transactions.
        
        In light of prior studies emphasizing the importance of secure and privacy-preserving REC transactions within blockchain networks, our research stands out through a unique fusion of privacy preservation and cutting-edge technology. Notably, we differentiate ourselves from previous works by placing a strong emphasis on privacy protection. This is achieved through the implementation of advanced cryptographic techniques, primarily zero-knowledge proofs. These proofs facilitate REC transaction validation while upholding the confidentiality of sensitive data, thereby elevating the security of REC trading. Furthermore, we introduce a novel dimension  by integrating Directed Acyclic Graph (DAG) technology alongside reputation-based consensus mechanisms. This integration notably bolsters the security and efficiency of our blockchain system. The efficacy of this innovative approach is substantiated through simulations, underscoring its potential to mitigate front-running attacks and enhance the overall performance of our blockchain technology.

\section{System Description}
\label{sec:model}
\begin{figure}
	\centering
	\includegraphics[width=\linewidth]{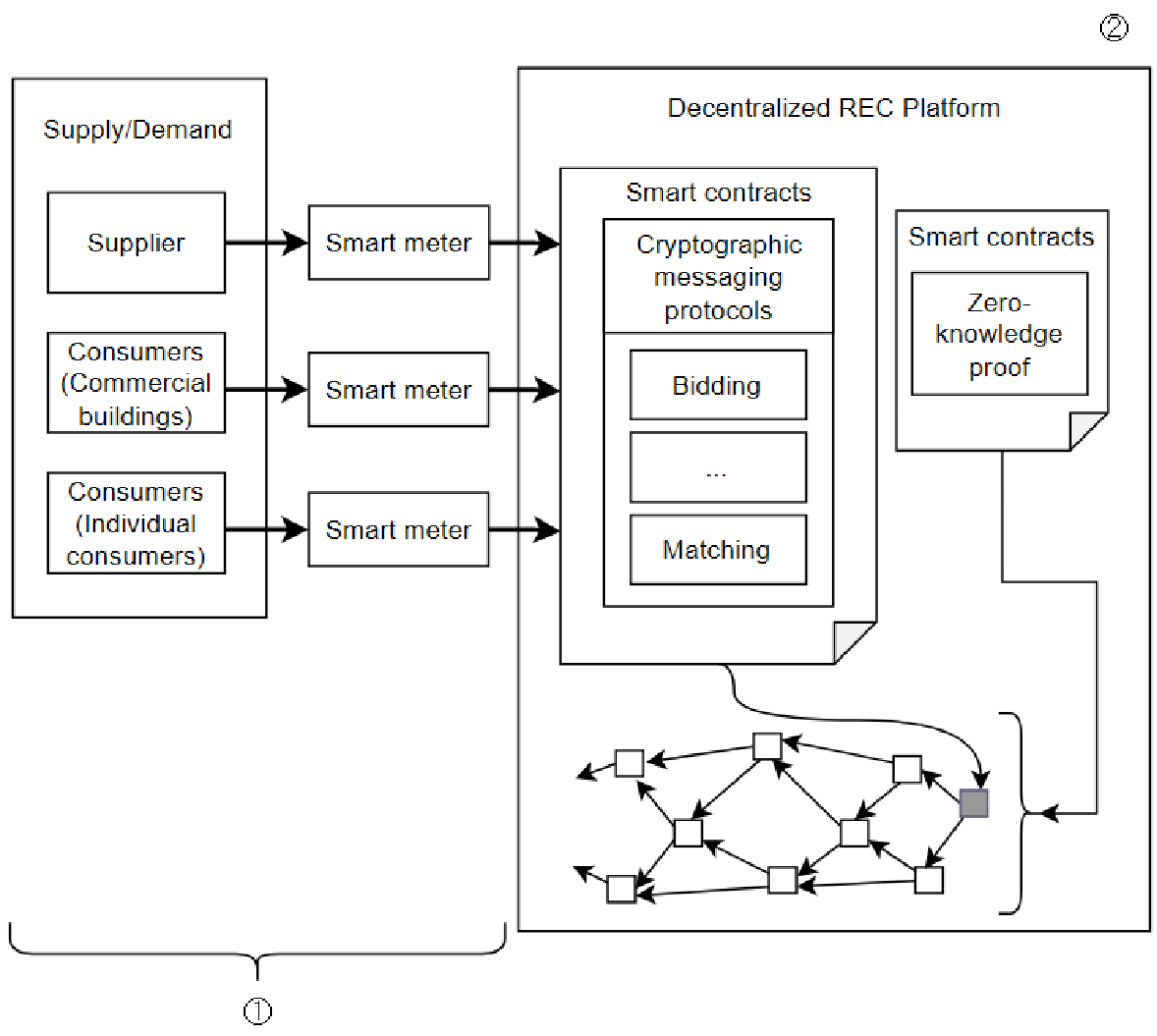}\\
	\caption{Renewable energy certificate trading system.}
	\label{fig:model}
\end{figure}

Fig.~\ref{fig:model} illustrates the REC trading platform considered in this study. Generally, REC trading consists of REC suppliers and commercial building consumers. Participants in the trading platform are equipped with a smart meter (Fig.~\ref{fig:model}, \raisebox{.5pt}{\textcircled{\raisebox{-.9pt} {1}}}) that records energy usage.

This study focuses on a blockchain-based platform (Fig.~\ref{fig:model}, \raisebox{.5pt}{\textcircled{\raisebox{-.9pt} {2}}}) to assist REC trading. The REC price fluctuates by supply-demand amount, and the demand will be incentivised in order to achieve the government regulation \cite{12baris}. The blockchain provides a tamper-proof mechanism, and a smart contract automatically trades for suppliers and consumers. Each REC seller and buyer can create transactions freely through cryptographic messaging protocols for their privacy protection. When a transaction is completed, the transaction record is placed into a pool until the pool size is around 1MB before it is packed into a new block on the blockchain. The main goal of the decentralized REC platform is to optimize trading performance and reduce human resources.

\subsection{Supplier Model}

Suppliers produce renewable energy, typically generated by photovoltaic (PV) panels, wind turbines, or biomass, which is further certified by the government or a certain organization. Consider that $i$th supplier generated renewable energy surplus in time slot $h$ as $E^{\text{res}}_{i, h}$, and one REC represents the renewable energy of one megawatt-hour. Hence, the REC generation by supplier $i$ can be formulated as follows:
\begin{equation} \label{eq:REC_generation}
	R^{\text{gen}}_{i,h} = \lfloor \frac{E^{\text{res}}_{i, h}}{1\text{ MWh}} \rfloor \lowperiod
\end{equation}
where $\lfloor \cdot \rfloor$ represents  the greatest integer less than or equal to the argument.

In the assumption, as governments and businesses prioritize renewable energy, the demand for Renewable Energy Certificates (RECs) will increase. This leads us to posit a linear relationship between the price of RECs and time, suggesting that the price of RECs will steadily decrease over time. 

Suppliers can use our trading platform to submit their RECs and sell them to consumers. Each supplier, denoted as $i$, sets their own selling price policy, represented by $P_i$. The selling price varies based on the source type and lifetime of the RECs. The total lifetime of the RECs is denoted as $\tau^{\text{lifetime}}$, while the remaining lifetime is expressed as $\tau^\text{remain}$. 

To formalize the selling price policy of supplier $i$, we use the following equation:

\begin{equation} \label{eq:suppliers_policy}
 P_{i,h}(b_{i,h}) = b_{i,h} \times \frac{\tau_\text{remain}}{\tau_\text{lifetime}} \lowperiod
\end{equation}
This equation shows that the selling price of supplier $i$ at time $h$ is determined by the bid amount $b_{i,h}$ and the remaining lifetime of the RECs. It is worth mentioning that the real-world relationship between time and pricing policies for REC instruments may be more complex and nonlinear; the linear assumption simply provides an initial conceptual framework that can be further refined.

\subsection{Consumer Model}

Commercial building consumers have energy storage systems, renewable energy sources, and smart meters. These consumers have specific renewable energy demands to meet the government's regulation in a year, i.e., a specific green ratio target which is determined by multiplying year's electricity consumption by a certain percentage from the government policy to reach. 

Suppose that there are $J$ commercial building consumers taking part in the REC trading market, and each of them has a green ratio represented by $G_j$. To encourage these consumers to participate in the market, the green ratio is measured once per hour and called $G_{j, h}$. The green ratio during a specific time slot $h$ is determined by the percentage of renewable energy to total energy consumption $E^{\text{c}}_{j, h}$ at that time, expressed as follows:

\begin{equation} \label{eq:Green_ratio}
 G_{j, h} = \frac{R^{\text{own}}_{j, h} \times {1\text{ MWh}} + E^{\text{res}}_{j, h}}{E^{\text{c}}_{j, h}}
\end{equation}

In this equation, $R^{\text{own}}_{j, h}$ is the number of RECs owned by consumer $j$, and $E^{\text{res}}_{j, h}$ is the amount of renewable electricity it generates or buys from the main utility.

In formulating our green ratio model, we initially excluded considerations for Energy Management Systems (EMS) due to the current assumptions. Acknowledging the insufficiency of current renewable energy levels to meet actual demands, we refrained from delving into this aspect in our present study, identifying it as a potential avenue for future research. However, we thoroughly incorporated factors such as renewable energy certificates, time constraints on renewable energy, and the annual target for achieving the green ratio \cite{22optimal}, deeming time a critical parameter in our model. 
      
Our model hinges on \eqref{eq:Green_ratio}, assuming that the government defines the green ratio (G) as the proportion of renewable energy to total energy consumption, mandating a specific percentage of renewable energy within the overall energy mix. This assumption aligns with the prevalent trend of governments setting targets to augment the utilization of renewable energy. Equation 3 is the concept that commercial buildings strive to fulfill the stipulated green ratio (G) by optimizing their green energy consumption over time without incorporating renewable energy storage within these buildings.

Furthermore, an hourly green ratio lag is defined as $G^{\text{lag}}_{j, h}$, to represent the gap between the green ratio target $G^{\text{target}}_j$ and the current green ratio $G_{j, h}$.
For consumer $j$, the green ratio lag $G^{\text{lag}}_{j, h}$ is expressed as:
\begin{equation}
	G^{\text{lag}}_{j, h} = \max(0, G^{\text{target}}_{j} - G_{j, h})
\end{equation}
When the renewable energy generated by commercial building itself cannot reach the green ratio within a particular period, there is a penalty that the building owner must pay. Therefore commercial building consumers need to buy REC in the trading market. 

Let the REC bidding policy of commercial building consumer $j$ be denoted as $C_{j, h}$. The bidding policy is affected by the green ratio lag $G^{\text{lag}}_{j, h}$. The bidding policy allocates a realistic bidding mechanism \cite{04Levin} with some adjustment; therefore, the bidding of commercial building consumers is as follows:
\begin{equation}\label{eq:building_policy}
	C_{j, h}(b_{j, h}) = b_{j, h} \times Q_{j, h} + D_{j,h} (\alpha_{j,h} G^{\text{lag}}_{j, h} \times E^c_h + Q_{j, h})^2 \lowperiod
\end{equation}
where $b_{j, h}$ represents the initial purchase price and $Q_{j, h}$ is the quantity to buy and is determined by $b_{j, h}$. Furthermore, there is a penalty when users cannot reach the yearly green target ratio; thus, the policy must include a penalty fee and time urgency. 
The penalty fee $D_{j,h}$ is structured in accordance with the regulations of different countries\cite{11haas}; it is proportional to the shortfall in meeting the standards, denoted as $G^{\text{lag}}_{j,h}$. 
 The penalty fee $D_{j,h}$ can thus be expressed as:
      \begin{equation}
          D_{j,h} = G^{\text{lag}}_{j,h}\times \gamma
      \end{equation}
      where $\gamma$ represents the coefficient associated with unmet capacity $G^{\text{lag}}_{j,h}$.
      
The time urgency of green ratio lag is modeled in the form of a quadratic equation\cite{14Moradi, 15Mahmoodi}. Because the purchase policy is subject to time constraints and green lags, a factor $\alpha_{j,h}$ representing the urgency of purchase by consumer $j$ in time slot $h$ is expressed as:
\begin{equation}\label{eq:building_policy_alpha}
	\alpha_{j,h} = \frac{t^{\max}}{t^{\text{remain}}} \times \frac{G^{\text{target}}_{j}}{G_{j, h}}
\end{equation}
where $t^{\text{max}}$ is the max time period to buy RECs and $t^{\text{remain}}$ represents the remaining time.

To prevent users from buying many RECs simultaneously and causing a monopolistic market \cite{70Fama}, we set up a limitation amount for consumer $j$ when bidding RECs, denoted as $Q^{\max}_j$. Let $p^{\max}$ represent the maximum REC selling price in the current market. Then, the maximum quantity for consumer $j$ permitted to buy in time $h$ can be written as:
\begin{equation}\label{eq:building_policy_quantity}
	Q_{j, h} = (1 - \frac{p^{\max}_h - b_{j, h}}{p^{\max}_h}) \times {Q^{\max}_{j, h}}
\end{equation}

If the bidding price $b_j$ is higher than the penalty fee $D_{j, h}$, the commercial building consumer can rather pay the penalty. In other words, only when the bidding price is lower than penalty, the commercial building buy RECs. Therefore, the desired quantity for consumer $j$ to buy is expressed as:
\begin{equation}\label{eq:building_quantity_constrain}
\begin{split}
	Q_{j, h} = \max(0, \min((1 - \frac{p^{\max}_h - b_{j, h}}{p^{\max}_h}) \times {Q^{\max}_{j, h}}, \\
	G^{\text{target}}_{j,h} \times E^c_h \times \gamma - D_{j, h}))
\end{split}
\end{equation}

Based on \eqref{eq:building_policy} and \eqref{eq:building_policy_quantity}, we can reformulate the commercial building consumer policy for buying RECs as:
\begin{equation}\label{eq:building_policy_updated}
	\begin{split}
		C_{j, h}(b_{j, h}) = {} & b_{j, h} \times (1 - \frac{p^{\max}_h - b_{j, h}}{p^{\max}_h}) \times Q^{\max}_{j, h} \\
		& + D_j(\alpha_{j,h} G^{\text{lag}}_{j, h} \times E^c_h + (1 - \frac{p^{\max}_h - b_{j, h}}{p^{\max}_h}) \\
		& \times Q^{\max}_{j, h})^2 \lowperiod
	\end{split}
\end{equation}

\subsection{Trading Model}

The blockchain platform matches orders between supply and demand participants through smart contracts. The demand side aspires to obtain RECs to reach the green ratio for its yearly requirements and buy RECs from the suppliers; the supply side generates and provides RECs and earns a profit. The selling quantity of supplier $i$ is $R_{i, h}$, and $Q_{j, h}$ is the RECs demand from commercial building consumers $j$ in time slot $h$. The quantity of RECs to buy in the market, $(Q_{j, h}, j = 1, 2, ..., J)$, and the quantity to sell, $(R_{i, h}, i = 1, 2, ..., I)$, in each time slot $h$ are determined as a supply-demand balance market, which can be further expressed as:
\begin{equation}\label{eq:trading}
	\sum_{j=1}^{J} Q_{j, h} = {\sum_{i=1}^{I} R_{i, h}} \lowperiod
\end{equation}

Consider that supplier $i$ can hold their RECs or sell RECs for a better price and thereby maximize their profits as depicted in \cite{04Levin}. The order-matching system matches the transaction from the highest REC price to the lowest, e.g., using a profit maximization model \cite{16Kumar}:
\begin{equation}
	\begin{split}
		\max_{b_{j, h}, p_{i, h}}  & \quad \sum_{j=1}^{J} b_{j, h}Q_{j, h} - \sum_{i=1}^{I} p_{i, h}R_{i, h} \\
		\text{subject to}
		& \quad b^{\text{min}}_j  \leq b_{j,h} \leq b^{\text{max}}_j \\
		& \quad p^{\text{min}}_i  \leq p_{i,h} \leq p^{\text{max}}_i
	\end{split}
\end{equation}
where $b^{\text{min}}$ and $b^{\text{max}}$ are denoted as the minimum and maximum consumer purchase prices; $p^{\text{min}}$ and $p^{\text{max}}$ express the minimum and maximum supplier selling prices, respectively.

\section{System Approach}\label{sec:approach}

The REC trading process uses the proposed blockchain, including smart contracts and the consensus mechanism. We consider two types of roles in the RECs trading scenario. The first is the renewable energy generators, who represent the RECs sellers, and the second is commercial building consumers, who represent the buyers. The REC trading follows the steps below to automate the REC transactions with privacy considerations through the blockchain: 1) For the data layer, the trading platform receives information from the seller about the REC price, source, number of RECs, created date, expiration date, and the generator's information. The information is organized and stored in the platform database; the blockchain only stores the transaction record. 2) For the network layer, the trading platform matches buyers and sellers and records the REC transaction information on the blockchain. 3) Transactions are recorded in the blockchain after a successful trade. However, to ensure the reliability of the transaction data, each new block must reach consensus at the consensus layer before it is packaged into the blockchain. The proposed consensus algorithm uses each user's reputation to determine the weight of the validated blocks, which makes this consensus algorithm faster and more robust than other methods.

\subsection{Data layer of REC trading}

Potential REC buyers search for RECs in the market and then start a cryptographic messaging protocol for sellers. After the transaction and payment are complete, the buyers can decrypt the REC data stored in the blockchain. The transaction records will be in the blockchain permanently after consensus validation.

The REC information and transaction records are essential and are placed into a pool until the pool size is around 1MB before it is packed into a new block on the blockchain. Table \ref{tab:tx_info} shows the data format to be recorded in a new block. Meanwhile, we apply the decentralized identifier (DID), defined by the World Wide Web Consortium \cite{20reed}. The DID is an identifier that enables a verifiable decentralized digital identity system to verify the identity and the owner's digital footprint. Moreover, like other conventional blockchain systems, it provides a traceable and tamper-proof feature for decentralized transactions by packaging transaction records and links in the blockchain.

\begin{table}[tph]
	\centering
	\caption{Transaction information}
	\label{tab:tx_info}
	\begin{tabular}{|c|c|}
		\hline
		\multicolumn{2}{|c|}{Transaction ID}  \\
		\hline
		Seller's DID    & Timestamp  \\ 	
		\hline
		Expiration date & REC Source \\
		\hline
		REC Price       & REC amount \\
		\hline
	\end{tabular}
\end{table}

\begin{figure}[tph]
	\centering
	\includegraphics[width=6cm]{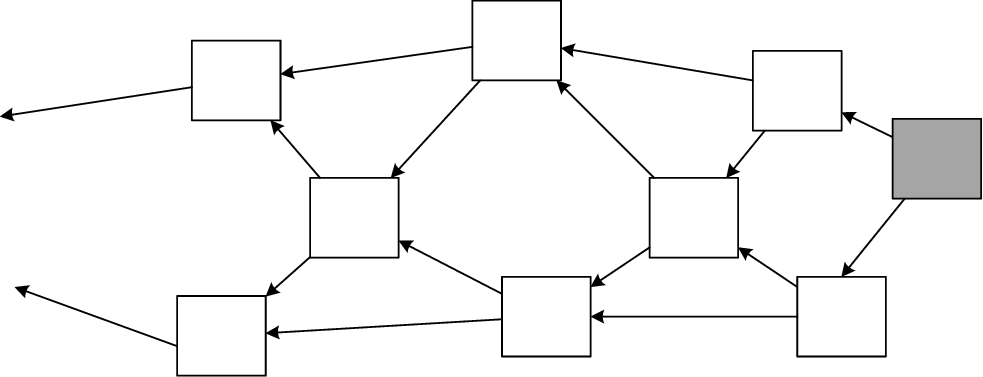}\\
	\caption{DAG-based blockchain}
	\label{fig:tangle}
\end{figure}

Fig.~\ref{fig:tangle} illustrates a DAG-based blockchain. Each node communicates with neighbor nodes and is verified by the two closest nodes. When a new block is added, two validations from other nodes are needed to detect tampering with transaction records and the REC information, which provides tamper-proof functionality. This mechanism reduces the validation time to create a new node and makes it easier to scale the blockchain ledger size.

\subsection{Network layer of REC trading}

Leveraging the benefit of DAG-based blockchain, we propose a trading platform and deploy the smart contract on it. Algorithm~\ref{alg:smart_contract} presents the pseudocode of the smart contract for REC trading. The smart contract can be divided into three major phases, pre-bidding (line 1--4), trading (line 5), and billing (line 6). In the pre-bidding phase, the buyer and seller must register for the platform and generate corresponding DIDs. During the trading phase, after the buyer and seller agree on the REC price, the transaction record is stored in the blockchain and verified by the blockchain consensus algorithm. After the bidding process is complete, the smart contract starts the billing phase and completes the token transaction. The use of $\gets$ represented as get value from the right hand side and $\Rightarrow$ expressed as message from left hand side to right hand side.

\begin{algorithm}
	\caption{Main Smart Contract}
	\label{alg:smart_contract}
	\begin{algorithmic}
		\State 1: BuyerList = []
		\State 2: function \textbf{DID}(User): \\
		\quad \! Generate user's public key $p_i$ and private key $q_i$ \\
		\quad \! Create DID document by using $p_i$ and $q_i$
		\State 3: function \textbf{seller}(Seller, REC):\\
		\quad \quad Store Seller.DID, $b^S_i$, $REC_q$, $t_{gen}$, $t_{exp}$
		\State 4: function \textbf{buyer}(Buyer): \\
		\quad \quad Store Buyer.DID in BuyerList
		\State 5: function \textbf{trading}(Seller, Buyer): \\
		\quad \quad sellerOTP $\gets$ Chaskey(Seller.DID, timestamp) \\
		\quad \quad buyerOTP $\gets$ Chaskey(Buyer.DID, timestamp) \\
		\quad \quad msgChannel $\gets$ Seller.Sign(timestamp, sellerOTP) \\
		\quad \quad $\hat{b}$ $\gets$ Buyer's purchasing policy \\
		\quad \quad i $\gets$ 0 \\
		\quad \quad for Buyer in BuyerList \\
		\quad \quad \quad Seller.msg(msgChannel) $\Rightarrow$ Buyer \\
		\quad \quad \quad Buyer.msg(buyerOTP, $\hat{b}$ + i) $\Rightarrow$ msgChannel \\
		\quad \quad \quad i $\gets$ i + 1 \\
		\quad \quad \quad if COMP($b^S_i$, $b^B_i$) = 1 \textbf{then} \\
		\quad \quad \quad \quad Store $S^{\text{trx}}$ \\
		\quad \quad \quad \quad  break
		\State 6: function \textbf{billing}(Seller, Buyer, $S^{\text{trx}}$) \\
		\quad \quad tokenReq $\gets$ Seller.Sign(Seller.DID, $S^{\text{trx}}$) \\
		\quad \quad Seller.msg(tokenReq) $\Rightarrow$ Buyer \\
		\quad \quad Buyer.validate(tokenReq, Seller.$p_i$) \\
		\quad \quad Buyer.msg($b^B_i$) $\Rightarrow$ Seller \\
		\quad \quad if tokenReq.isFinish() then \\
		\quad \quad \quad REC.owner = Buyer
	\end{algorithmic}
\end{algorithm}

The seller submits the REC encrypted with the private keys to the trading platform in the pre-bidding phase. The duties of the platform are to check the integrity of potentially forged RECs. The smart contract matches the order and records the transaction on the blockchain after the REC is submitted and encrypted during the trading process. After the seller signs the REC with their private key $q^U_i$, the trading platform stores the bidding price $b_i$. The REC information $S^{\text{REC}}$ is encrypted by the SHA256 encryption algorithm \cite{12kasgar}. The encryption obtained by information of $S^{\text{REC}}$ with private key $q^U_i$ is denoted as $\text{Sign}(e(S^{\text{REC}}, q^U_i))$ and the REC to be sold can be obtained as:
\begin{equation}
	REC_q = \text{Sign}(e(S^{\text{REC}}, q^U_i) ,t_{\text{gen}},t_{\text{exp}}) \lowcomma
\end{equation}
where Sign($\cdot$) is the digital signature function that is obtained by decrypting $\text{Sign}(e(S^{\text{REC}}, q^U_i))$ with public key $p^U_i$ and is denoted as $\text{Sign}((e(S^{\text{REC}}, q^U_i)), p^U_i)$. $t_{\text{gen}}$ is the time the REC was generated, and $t_{\text{exp}}$ is the certificate's validity duration.

\subsection{Consensus layer of REC trading with privacy preserving}

To reduce the complex computation of a typical consensus, we apply FPC instead. When a new transaction is created, it goes through multiple voting mechanisms to complete the consensus algorithm. This mechanism confirms whether this currently added transaction is valid and should be placed in the blockchain. A forged or invalid transaction will not be placed in the blockchain. The voting mechanism is a decentralized verification mechanism where each node represents a voting right and cooperates to complete the verification and ensure the robustness of the blockchain network.

During the consensus process, the selected blockchain nodes which can vote are denoted as $n\in \mathbf{\mathcal{N}}$, and the number of voting rounds is denoted as $k\in \mathbf{\mathcal{K}}$. To clarify the mechanism of the consensus algorithm, let us suppose there are two new nodes denoted by $n_1$ and $n_2$. When these two nodes independently declare that each is the only valid transaction at the moment, the transactions are in conflict. Therefore, a validation mechanism is activated to determine which one is correct. This mechanism starts with a randomly selected a threshold value from 0 to 1, denoted as $\omega$, which satisfies $0 \leq \omega \leq 1$. At each round the other nodes near by $n_1$ and $n_2$ are subsequently interrogated $k$ times. In the first round of voting, the threshold $\omega$ is drawn from a system-defined value and is limited to $0.5 \le \omega \le 1$. From the second round to the final round, the threshold $\omega$ is taken from $[\beta, 1-\beta]$, where $\beta$ is a random value with $0 < \beta < 0.5$. After $\mathbf{\mathcal{K}}$ rounds of voting, if the proportion of $n_1$ is greater than $\omega$, the tentative opinion of the node is changed to $n_1$. Otherwise, it is changed to $n_2$. Moreover, if $n_1$ is validated successfully, $n_2$ is isolated and can never perform the consensus process.

Algorithm~\ref{alg:consensus} presents the proposed FPC with an integrated reputation mechanism as our consensus algorithm. The consensus algorithm is based on voting results from nearby nodes, and the voting weight of each verifier is adjusted proportionally to its reputation. 

During the consensus process, the higher the reputation value of nodes, the higher the probability of participating. Each node in the network must increase its reputation value by participating in transactions. After each transaction, the consensus mechanism gives a reputation value to the participants. Let $r^\text{activity}_{i, k}$ denote the rating given by node $i$ in the $k$th transaction and let $\lambda$ be a system-designed decay constant. $r^\text{activity}_{i, k} = 0$ means that the former participant in the $k$th interaction is inactive, and $r^\text{activity}_{i, k} = 1$ means that the former participant in the $k$th interaction is active. The reputation value decreases over time. We use $t^\text{c}$ to represent the current time and $t^\text{l}$ to represent the last active time. The $\delta$ is the time difference between the current and last active times. Thus, the reputation of node $i$ can be updated as:
\begin{equation}\label{eq:reputation_decay}
	\begin{split}
		& \delta \gets t^\text{c} - t^\text{l} \\
		& r^\text{activity}_{i, k} \gets r^\text{activity}_{i} \times \exp(-\lambda \times \delta)
	\end{split}
\end{equation}
The voting weight is denoted as $r^\text{weight}_i$ calculated by dividing $r^\text{activity}_{i, k}$ by the reputation of all nodes $\sum^N_{i=1} {r_{i, k}}$ in the voting process:
\begin{equation}\label{eq:reputation_wwight}
	r^\text{weight}_ i = r^\text{activity}_{i, k}  / \sum^N_{i=1} {r_{i, k}}
\end{equation}
The verifier can validate a new transaction by voting:
\begin{equation}\label{eq:opinionQuery}
	\begin{split}
		\text{opinionQuery}& \gets \text{response}[S^{\text{trx}}_{\text{index}}, n] \times r^\text{weight}_{i} \\
		\text{Voting Result}& {} =
		\begin{cases}
			\text{opinionQuery}, & \text{if } \text{opinionQuery} \geq \omega \\
			0,                   & \text{otherwise}                           
		\end{cases}	
	\end{split}
\end{equation}

In \eqref{eq:opinionQuery}, the verifiers resolve conflicts and correct the received transactions by their reputation weights. opinionQuery($\cdot$) is a function that is used for asking the other verifier nodes to verify transaction $S^{\text{trx}}_{\text{index}}$ and determine the opinion with its reputation. If the node has a higher reputation, where a greater weight verifies that the transaction is valid; however, the voting result will be ignored when the reputation value is too small. If the total number of votes surpasses a threshold, all verifiers successfully validate the transaction and reach a consensus.

\begin{algorithm}
	\caption{Fast Probability Consensus with Reputaion}
	\label{alg:consensus}
	\begin{algorithmic}
		\Require voting rounds $\mathbf{\mathcal{K}}$; nodes $\mathbf{\mathcal{N}}$ 
		\State 1: \textbf{Loop} for each round $k$
		\State 2: \quad \textbf{If} $k$ == 1 \textbf{then}
		\State 3: \quad \quad  $\omega \gets$ random(a, b) where $0.5 \le a \leq b \le 1$
		\State 4: \quad \textbf{Else}
		\State 5: \quad \quad $\omega \gets$  random($\beta$, 1 - $\beta$) where $0 \le \beta \leq 0.5$
		\State 6: \quad \textbf{For} $n$ in $\mathbf{\mathcal{N}}$ \textbf{do}
		\State 7: \quad \quad Calculate $r^\text{activity}_{i, k}$ using \eqref{eq:reputation_decay}
		\State 8: \quad \quad Calculate $r^\text{weight}_{i}$ by \eqref{eq:reputation_wwight}
		\State \quad \quad \quad /* Ask near node voting option */
		\State 9: \quad \quad opinionQuery[$S^{\text{trx}}_{\text{index}}$, $n$] = respond[$S^{\text{trx}}_{\text{index}}$, $n$] $\times$ \\
		\quad \quad \quad $r^\text{weight}_{i}$
		\State 10: Calculate the avg(opinionQuery)
		\State 11: \textbf{If} avg(opinionQuery) $\geq \omega$ \textbf{then}
		\State 12: \quad \textbf{Return} validate and isolate the confict node
		\State 13: \textbf{Else}
		\State 14: \quad \textbf{Return} invalidate and isolate current node
	\end{algorithmic}
\end{algorithm}

In a blockchain network attack, a single entity can create multiple fake identities, or "Sybils," to gain an unfair advantage in control of a network or manipulate its behavior. This can be a major issue in decentralized networks, where the identities of nodes are not necessarily known or trusted. The specific methods used to defend against Sybil attacks can vary depending on the system but generally involve some combination of authentication and reputation systems.

In this research, the REC supplier is asked to be certificated that they can generate renewable energy and sell it as a REC. Hence, the authentication method is applied for the supplier side to defend against the Sybil attack. On the other hand, the proposed consensus embedded with a reputation in the voting mechanism, an initial reputation value is given first and can be decreased by time or less active or increased after successful verification. Furthermore, when a node has been found faking a verification, the nodes will be disconnected from the network.

The proposed method strongly emphasizes authenticated suppliers and consumers, utilizing their reputations as voting weight concepts. However, in terms of security considerations, it is important to acknowledge that the proposed approach, relying on reputation as a basis for validation, may not equate to the robustness of hashing power-based Proof of Work (PoW) consensus mechanisms. PoW excels in scenarios involving fully untrusted nodes, providing high security within the network. The tradeoff, however, comes in the form of substantial computational power required for hashing calculations and extended verification times for processing transactions and incorporating them into the blockchain.

The reputation-based consensus utilized in this study introduces a different security tradeoff. While it may not possess the same level of security as PoW, it fosters trust and reliability among participants with established reputations. It shifts the security focus towards maintaining the integrity of participants' reputations and encouraging honest behavior within the network. Striking a balance between security, trust, and efficiency is pivotal in blockchain design, and the proposed method achieves this balance by leveraging reputation as a vital aspect of the consensus algorithm. 

It is essential to recognize that the choice of consensus mechanism is a deliberate design consideration influenced by the specific use case and desired network dynamics. The proposed approach prioritizes decentralization and validation randomness conducive to the context of renewable energy trading. By carefully tailoring the consensus mechanism to suit the objectives of the renewable energy trading ecosystem, we ensure that the network remains secure, efficient, and reliable for all participating entities.

Algorithm \ref{alg:privacy} presents the pseudocode for privacy-preservation in the trading phase. At the beginning of trading, the algorithm calculates the user's weight. The weight is used to determine whether to create a new proxy account by looking at the user's activity and transaction volume, which is positively correlated to the user's transaction volume and activity status.

According to the different weights of the users, the trading process can be separated into three different cases. In the first case, trading is done directly through the original account when the activity status is lower than average. Otherwise, an account creation mechanism is established, it can be further subdivided into a proxy account or an empty account. As the name implies, proxy accounts are a middleman for the counterparty linked to the original seller's identity. They are created when the volume of transactions is large. In addition, multiple proxy accounts are created and spread evenly. On the other hand, empty accounts are temporary accounts for small-volume transactions and are not directly linked to the original account.

\begin{algorithm}
	\caption{Privacy Preservation Mechanism for Blockchain trading}
	\label{alg:privacy}
	\begin{algorithmic}
		\Require transaction $S^{\text{trx}}_{\text{index}}$; trading nodes $\mathbf{\mathcal{N}}$ \\
		Initialize average amount of nodes in time slot $h$ $U_\text{Amount}$; average activity of nodes in time slot $h$ $U_\text{Activity}$
		\State 1: \textbf{For} each node in $\mathbf{\mathcal{N}}$ \textbf{do}
		\State 2: \quad \textbf{If} $S^{\text{trx}}_{\text{index}}$ amount $\leq$ $U_\text{Amount}$ \textbf{then}
		\State 3: \quad \quad \textbf{return} current account
		\State 4: \quad \textbf{Else if} $S^{\text{trx}}_{\text{index}}$ amount $>$ $U_\text{Amount}$  \textbf{then}
		\State 5: \quad \quad \textbf{If} $S^{\text{trx}}_{\text{index}}$ activity $\leq$ $U_\text{Activity}$
		\State 6: \quad \quad \quad \textbf{return} new empty account
		\State 7:\quad \quad \textbf{End If}
		\State 8: \quad \quad \textbf{return} with proxy account
		\State 9: \quad \textbf{End If}
		\State 10: \textbf{End for}
		\State 11: Continue trading and records on the blockchain with \\
		\quad \text{} \text{} \text{} the return account.
	\end{algorithmic}
\end{algorithm}

\section{Numerical Results}\label{sec:result}

\begin{figure}
	\centering
	\includegraphics[width=\linewidth]{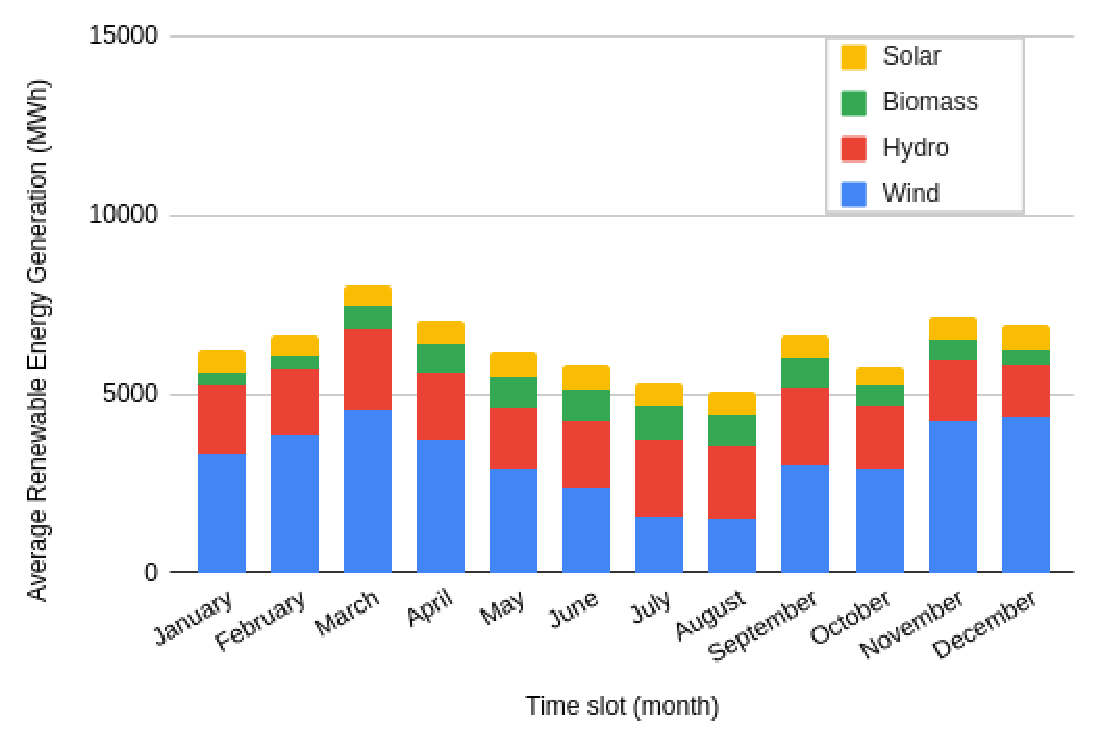}\\
	\caption{Renewable energy generation from PJM dataset}
	\label{fig:pjm_data}
\end{figure}

This study examined the effectiveness of the DAG integrated reputation mechanism consensus algorithm blockchain approach to REC trading. The study examined 15 commercial building consumers, and 100 nodes as validators in the blockchain network. The REC supplier, commercial buildings consumers were equipped with smart meters and connected to the self-host trading platform. In the simulation, we added a built-in script that would connect to the ethereum's API to buy or sell the RECs according to the commercial building consumer's green ratio target. When the platform received a trading request, the REC information was piped into the proposed blockchain schema and matched to the orders by the closest price. 

The energy demand for commercial building consumers was derived from the dataset in the PJM miner \cite{PJM} about energy demand during 2021. The supply side was simulated using dataset of four renewable energy sources from in the same year, which are shown in Fig.~\ref{fig:pjm_data}. The data was reported daily then organized to monthly to demonstrate the renewable generation resources in megawatts for several sources: biomass, hydroelectric, wind, and solar PV.

The green ratio was set according to different user types. If the user did not reach the green ratio target, a penalty fee must paid according to (5), and unmet capacity coefficient $\gamma=\$10000$/kWh were set\cite{mea-reda}. In the proposed consensus algorithm, each transaction node decayed after each time slot, using the decay parameter $\lambda = 0.2$. This decay enhanced user participation and ensured that users participating in the verification node were more trustworthy than users participating in lower-activity nodes. Furthermore, the system-defined threshold $\omega = 0.5$ was set \cite{19toward}.

\begin{figure}
	\centering
	\includegraphics[width=\linewidth]{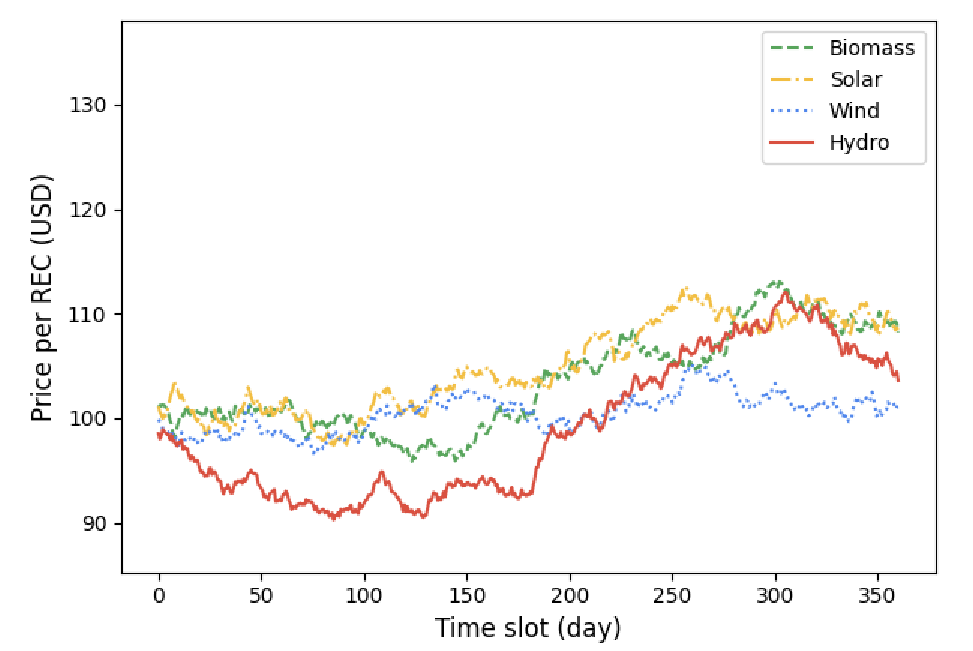}\\
	\caption{Market tradings of REC between 15 commercial building consumers consumers in one year}
	\label{fig:trading1}
\end{figure}

Fig.~\ref{fig:trading1} recorded the REC transaction price trend on the proposed blockchain platform in one year. The REC's initial bidding price started at US\$100 and grew higher due to the consumers' green ratio urgency and the scarcity of REC source types. The REC price was also affected by the lifetime limit and quantity generated. For example, because most renewable energies were generated by wind, the REC price for wind resources was relatively lower than for other resource types. The price dropped for the different source types of REC, owning to the lifetime limit.

\begin{figure}
	\centering
	\includegraphics[width=\linewidth]{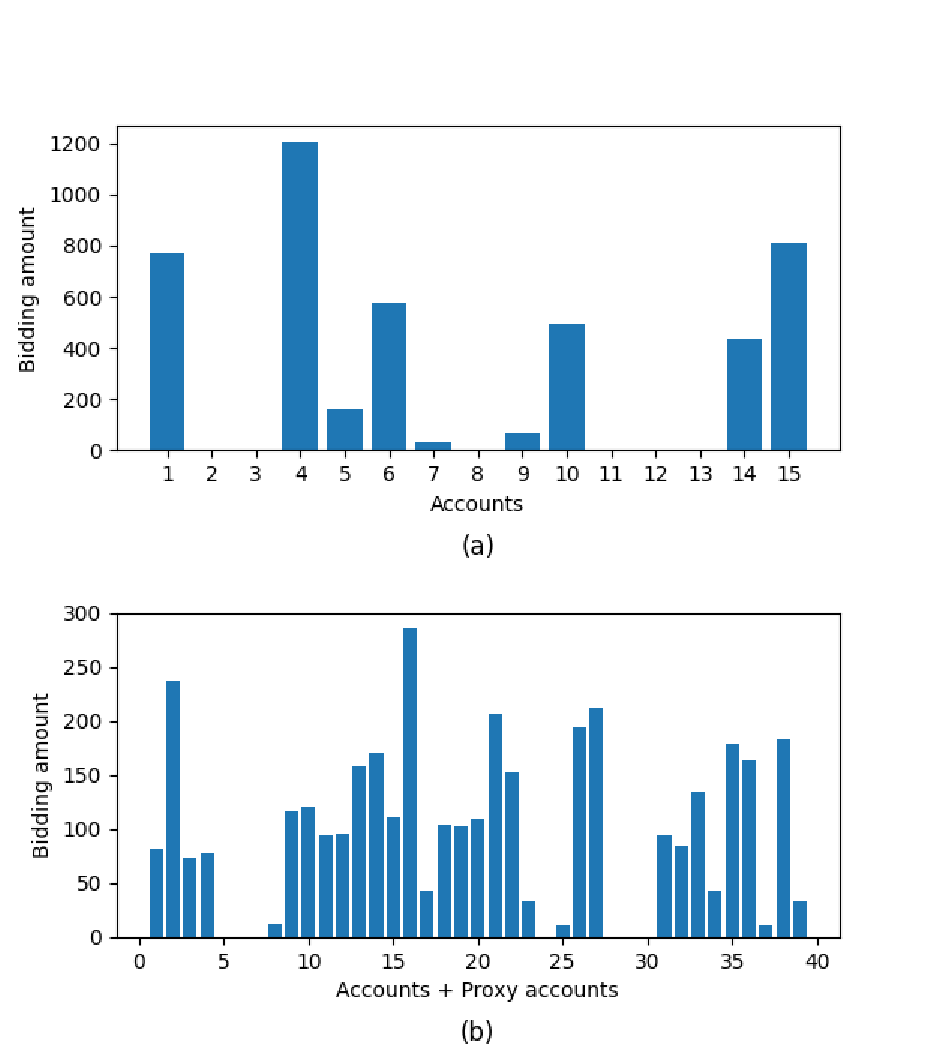}\\
	\caption{Trading distributions between (a) without generating proxy transaction accounts and (b) with generating proxy transaction accounts. When the transaction process does not generate a distribution of transactions for proxy accounts, each account is mapped to each consumer. A total of 15 consumers' trading quantities are recorded on the platform, and it can be guessed that the user is a important commercial building consumer. For example, account 4 was likely to be a commercial user due to the highest transaction quality. However, the figure below shows the privacy-preserving pattern by generating proxy transaction accounts during the transaction. The total number of accounts traded was the original 15 accounts plus the proxy accounts, and the total number of accounts was 40.}
	\label{fig:privacy}
\end{figure}

Fig.~\ref{fig:privacy} shows the original distribution of REC trading and the distribution after applying the proposed privacy-preserving method. Fig.~\ref{fig:privacy} (a) shows the trading distribution without generating proxy accounts. Each account trading on the platform is mapped to each consumer. A malicious attacker can analyze the trading quantity to know which accounts are important commercial building consumers. However, in Fig.~\ref{fig:privacy} (b), we applied a privacy-preserving method during the trading process by generating proxy transaction accounts and distributing the bidding more evenly than without it. Therefore, attackers have greater difficulty using a side-channel attack to analyze the users on the platform.

\begin{table}
	\centering
	\caption{Transaction Time of different consensus algorithms.}
	\begin{tabular}{|cccccc|}
		\hline
		\multicolumn{6}{|c|}{Transaction Time per REC}
		\\ \hline
		\multicolumn{1}{|p{2cm}|}{Blockchain Ledger Size} & \multicolumn{1}{|p{1cm}|}{Green ratio} & \multicolumn{1}{c|}{PoW \cite{99jakobsson}}       & \multicolumn{1}{c|}{Prism \cite{19bagaria}} & \multicolumn{1}{c|}{PoS \cite{21saleh}}     & \multicolumn{1}{|p{1cm}|}{Proposed Method} \\ \hline
		\multicolumn{1}{|c|}{\multirow{3}{*}{100}}        & \multicolumn{1}{c|}{30\%}        & \multicolumn{1}{c|}{0.00639s}  & \multicolumn{1}{c|}{0.00233s} & \multicolumn{1}{c|}{0.0004s} & 0.00053s \\ \cline{2-6}
		\multicolumn{1}{|c|}{}                            & \multicolumn{1}{c|}{60\%}        & \multicolumn{1}{c|}{0.00658s}  & \multicolumn{1}{c|}{0.00241s} & \multicolumn{1}{c|}{0.0004s} & 0.00054s \\ \cline{2-6}
		\multicolumn{1}{|c|}{}                            & \multicolumn{1}{c|}{90\%}        & \multicolumn{1}{c|}{0.00662s}  & \multicolumn{1}{c|}{0.00309s} & \multicolumn{1}{c|}{0.0005s} & 0.00053s \\ \hline
		\multicolumn{1}{|c|}{\multirow{3}{*}{1000}}       & \multicolumn{1}{c|}{30\%}        & \multicolumn{1}{c|}{5.3727s}   & \multicolumn{1}{c|}{1.8983s} & \multicolumn{1}{c|}{0.0022s} & 0.00153s \\ \cline{2-6}
		\multicolumn{1}{|c|}{}                            & \multicolumn{1}{c|}{60\%}        & \multicolumn{1}{c|}{5.7421s}   & \multicolumn{1}{c|}{2.1054s} & \multicolumn{1}{c|}{0.0027s} & 0.00163s \\ \cline{2-6}
		\multicolumn{1}{|c|}{}                            & \multicolumn{1}{c|}{90\%}        & \multicolumn{1}{c|}{6.1936s}   & \multicolumn{1}{c|}{2.9255s} & \multicolumn{1}{c|}{0.0029s} & 0.00153s \\ \hline
		\multicolumn{1}{|c|}{\multirow{3}{*}{10000}}      & \multicolumn{1}{c|}{30\%}        & \multicolumn{1}{c|}{521.3334s} & \multicolumn{1}{c|}{189.4180s} & \multicolumn{1}{c|}{0.0281s} & 0.00169s \\ \cline{2-6}
		\multicolumn{1}{|c|}{}                            & \multicolumn{1}{c|}{60\%}        & \multicolumn{1}{c|}{548.2847s} & \multicolumn{1}{c|}{239.2054s} & \multicolumn{1}{c|}{0.0355s} & 0.00209s \\ \cline{2-6}
		\multicolumn{1}{|c|}{}                            & \multicolumn{1}{c|}{90\%}        & \multicolumn{1}{c|}{561.3944s} & \multicolumn{1}{c|}{263.1162s} & \multicolumn{1}{c|}{0.0379s} & 0.00329s \\ \hline
	\end{tabular}
	\label{loginMock}
	\label{tab:tx_time}
\end{table}

Table~\ref{tab:tx_time} shows the results of REC transactions on the blockchain platform for different blockchain ledger sizes and green ratio targets. The ledger size means the total blocks in the blockchain network, and each block size is approximately 1MB. Because of each transaction's varying size, each block may contain unequal transaction records. For a fair comparison, each block size is fixed and only considers the relationship between transaction time and ledger size. Also, these consumers are asked to reach a specified green ratio within a year, the urgency of purchasing RECs upfront is more critical when the green ratio target is higher. The chosen green ratio targets of 30\%, 60\%, and 90\% indicate the potential trajectory of government policies and industry trends \cite{14pazheri}. These numbers could represent ideal environmental targets crucial for sustainability and environmental protection. For instance, 30\% signifies a fundamental sustainable development target, 60\% represents a more aggressive effort, and 90\% stands for a highly challenging yet achievable ultimate goal. Furthermore, intervals like 30, 60, and 90 may represent a gradual process of achieving higher green ratios. This progressive approach aids organizations and industries in transitioning gradually to meet more environmentally friendly objectives while mitigating risks and costs during the transformation. Due to the higher green ratio target, users need to transact more frequently with the platform, the blockchain supports more transactions that need to be packaged and verified, which affects the whole network's transaction speed. Because PoW refers to the proof that a certain amount of computational effort has been spent, PoW takes the longest time for each transaction. PoS refers to the selection of verifiers in proportion to the number of tokens staked; it does not require substantial computational power and has a shorter transaction time than PoW. Moreover, PoW consensus verification is slowest when the blockchain ledger size increases, as finding the nonce of new blocks becomes more challenging. Prism is a scaling solution based on PoW, it can be divided into multiple PoW chains and have a significant improved transaction time. PoS requires validators to hold and stake tokens for verification, yet still limited by the standard blockchain structure design and affect the transaction performance when the blockchain ledger size grows. In contrast, the proposed reputation-based FPC considers only the voting results of selected nodes and uses the network architecture of DAG, so the transaction speed is faster than PoW and PoS when the blockchain ledger size or green ratio increases.

\begin{figure}
	\centering
	\includegraphics[width=\linewidth]{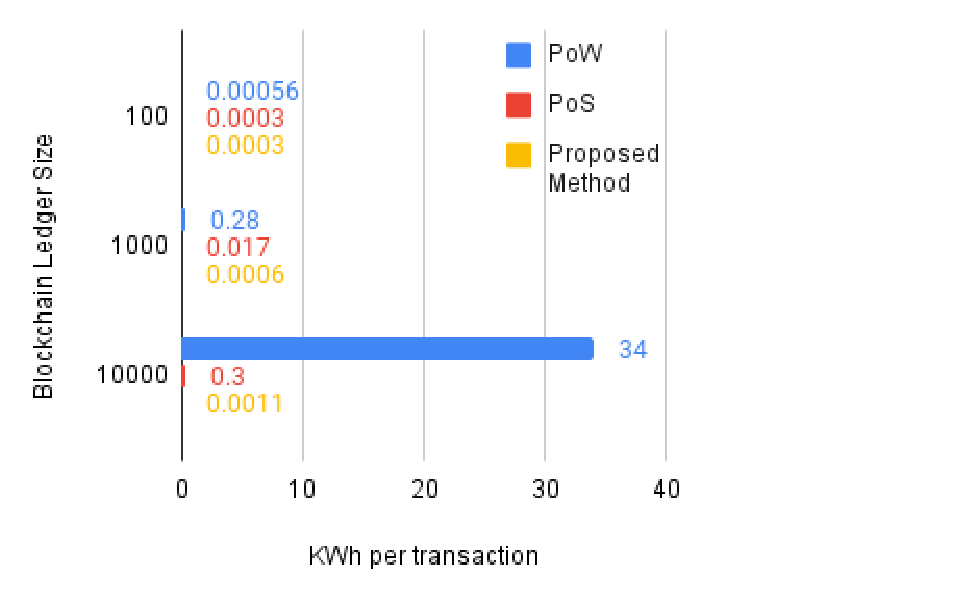}\\
	\caption{Transaction energy consumption of different consensus algorithms. As the blockchain ledger size increases, PoW is less likely to mine the next block, thus making PoW, which initially relies on a lot of computation power, consume more energy as the blockchain ledger size gets larger. PoS reduces the PoW computation power by replacing mining with staking tokens. On the other hand, the proposed method always has a voting mechanism through nearby nodes, thus keeping the energy consumption lower and less affected by the increase in blockchain ledger size.}
	\label{fig:tx_energy}
\end{figure}

Fig.~\ref{fig:tx_energy} shows the energy consumption of different consensus algorithms. PoW consumes more energy than the others because it has to calculate the nonce to generate the next new block. PoS eliminates the nonce calculation and replaces it with stake tokens; hence, the energy consumption for calculation is lower in each transaction. However, the standard blockchain structure design increases energy consumption as the blockchain ledger size increases. By contrast, the proposed reputation-based FPC maintains a low energy consumption and is less affected by the blockchain ledger size, so it has better scalability for the blockchain network.

\begin{figure}
	\centering
	\includegraphics[width=\linewidth]{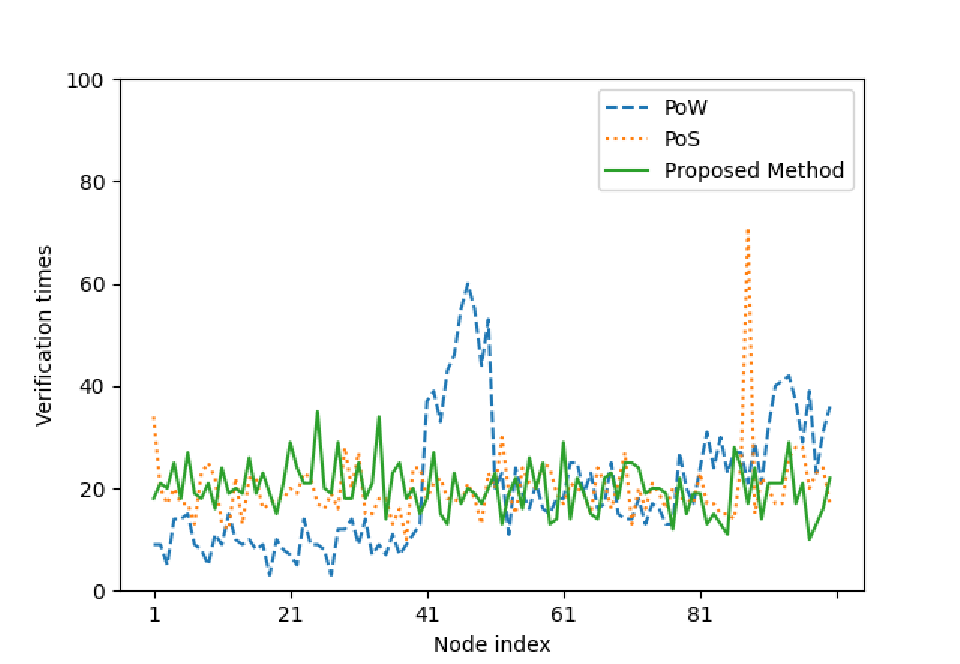}\\
	\caption{Verification times distribution between different consensus algorithms.}
	\label{fig:distribute}
\end{figure}

Because the blockchain is a distributed system maintained by the participants, its security relies on its consensus algorithm. PoW used the computational resources as validation weights, PoS used the number of tokens held by verifiers, and the proposed method verified by using the verifier's reputation as the voting weight. When one entity controls the majority of the votes, the network decision-making process tends to act for the benefit of a specific entity rather than the network. Therefore, maintaining a certain extent of decentralization and randomness of validation is key to blockchain consensus.

Fig.~\ref{fig:distribute} shows the distribution of the 3000 times of validations among 100 nodes for different consensus algorithms. The x-axis indicates the node index, and the y-axis shows the number of verification times for each node. Verification times refers to the number of times a node verifies a new transaction as a valid transaction. Therefore, the more times a node verifies, the more absolute verification power the node has. Using the standard deviation in statistics, when the standard deviation value is larger, it indicates that the degree of centralization in the overall network is higher, while a smaller standard deviation value indicates a higher degree of decentralization. From the Fig.~\ref{fig:distribute}, the PoW had the largest standard deviation, the PoS was the second placed, and the proposed method was the smallest of standard deviations. Their respective numerical standard deviations average were: 17.3 for PoW, 6.1 for PoS, and 4.4 for the proposed method in average 5 test results. Therefore, we can conclude that the proposed method has better decentralization.

Our method underwent a rigorous validation process to confirm its effectiveness and robustness. Initially, we conducted simulations for various green policy targets, including REC holding policies of 30\%, 60\%, and 90\%. These diverse objectives represent the spectrum of demands commercial building consumers might have, ranging from a low to a high green ratio.

Through these simulations, we quantified our method's performance under different conditions, demonstrating its adaptability to a wide range of green policy objectives. We found that our method consistently exhibited optimal results across varying policy objectives, affirming its stability and superiority in diverse contexts.
      
Furthermore, we systematically compared our method with commonly used consensus algorithms in controlled experiments, including Proof-of-Work (PoW) and Proof-of-Stake (PoS). These comparative tests further underscored the unique advantages of our method, particularly in terms of energy consumption and adaptability to changes in blockchain ledger size.
      
Our method demonstrated exceptional robustness through extensive testing and comparison with other approaches. These validations ensure that our method results from careful deliberation, comprehensive evaluation, and empirical study and maintains high efficiency across different conditions.
      
\section{Conclusion and Future Work}\label{sec_conclusion}

This paper investigates an efficient and trustless REC market platform. The REC market is composed of providers and consumers with smart meters with the addition of validators to help validate RECs. In the study, we investigated commercial consumers that bid for RECs based on energy types and price to reach their green energy ratios. We proposed a DAG-based blockchain system for the REC trading platform to address the active markets and trading efficiency. The proposed FPC-based consensus algorithm associated with reputation and privacy preservation was developed accordingly. The proposed approach allows REC suppliers and consumers to take advantage of blockchain benefits and avoid wasting human resources or energy for computation. Real-world data, including renewable energy generation and requirements, were used to evaluate performance. Our analysis showed that the consumers' privacy protection schema is effective for trading RECs on the blockchain. Also, the proposed reputation-based consensus was faster by 41\% and consumed less energy by 65\% than proof-of-stake. The future work will investigate the limitation of proof-of-reputation consensus that cannot be trustful when too many new consumers join to launch an attack and provide more robust security.

\section*{Acknowledgements}
This work was supported by the Ministry of Science and Technology of Taiwan under Grant MOST 110-2221-E-007-097-MY2.


\end{document}